\documentclass[aps,prb,twocolumn,showpacs,superscriptaddress,groupedaddress]{revtex4}
\usepackage{graphicx}  % needed for figures
\usepackage{dcolumn}   % needed for some tables
\usepackage{bm}        % for math
\usepackage{amssymb}   % for math
\usepackage{amsmath} 
\usepackage{natbib}
\usepackage{color}
\usepackage{ulem} % needed for \scrap-command

\definecolor{grey}{rgb}{.6,.6,.6}

\begin{document}

\title{Quantum Properties of the radiation emitted by a conductor in the Coulomb Blockade Regime}

\author{C. Mora}
\affiliation{Laboratoire Pierre Aigrain, \'Ecole Normale Sup\'erieure-PSL Research University, CNRS, Universit\'e Pierre et Marie Curie-Sorbonne Universit\'es, Universit\'e Paris Diderot-Sorbonne Paris Cit\'e, 24 rue Lhomond, 75231 Paris Cedex 05, France}
\author{C. Altimiras}
\author{P. Joyez}
\author{F. Portier}
\affiliation{Service de Physique de l'Etat Condens\'e, CEA, CNRS, Universit\'e Paris-Saclay, CEA Saclay, 91191 Gif-sur-Yvette, France}

\date{\today}

\begin{abstract}
We present an input-output formalism describing a tunnel junction strongly coupled to its electromagnetic environment. We exploit it in order to investigate the dynamics of the radiation being emitted and scattered by the junction. We find that the non-linearity imprinted in the electronic transport by a properly designed environment generates strongly squeezed radiation. Our results show that the interaction between a quantum conductor and electromagnetic fields can be exploited as a resource to design simple sources of non-classical radiation. 
\end{abstract}

\maketitle

\section{Introduction}

Circuit quantum electrodynamics (cQED) describes at a quantum level the interaction between  electromagnetic fields and artificial atoms implemented by quantum conductors such as Josephson junctions \cite{devoret2013superconducting} or quantum dots \cite{petersson2012circuit,FreyDpotCavityPRL2012,viennot2015coherent}. This new architecture has triggered a number of pioneering experiments \cite{HofheinzArbitraryStatesNature2009,KirchmairSinglePhotonKerrNature2013,murch2013reduction,BretheauDynamicZenoScience2015}. However quantum conductors can also be continuously driven out-of-equilibrium by dc biases, giving rise to situations having no evident counterpart in atomic physics. Recent predictions and experiments relevant to these situations already point to interesting quantum electrodynamics effects. To cite a few, dc driven quantum conductors can be used as stochastic amplifying media giving rise to lasing (or masing) transitions in the field stored in RF cavities \cite{Padurariu2012,ChenSCPTlaserPRB2014, LiuDQDmaserScience2015}, as sources of sub-poissonian \cite{beenakker2001,BeenakkerSchomerusantibunchedPhotonsQPCPRL2004,FulgaFtempQPCPRB2010,LebedevPhotonStatsQPCPRB2010,hassler2015} and squeezed radiation \cite{LeppakangasPhotonPairsPRL2013,armour2013,gramich2013,LeppäkangasNJOP2014,AnkerholdPhotonPairsPRB2015,trif2015,GasseTunnelSqueezePRL2013,ForguesTunnelPairsPRL2014,mendes2015,grimsmo2016}. Conversely, non-classical features of an incoming field may be revealed in the $I(V)$ curves of the conductor~\cite{Souquet2014}.

The common underlying mechanism for these effects is the probabilistic transfer of discrete charge carriers through the quantum conductors. The resulting current fluctuations excite the surrounding electromagnetic environment. This coupling not only results in photon emission, but also modifies the transport properties of the conductor itself \cite{delsing1989,geerligs1989,ClelandDCBinTunnelPRB1992,HolstDCBPRL1994, AltimirasDCBQPC2007,PekolaDynesDCBPRL2010,HofheinzBrightDCBPRL2011,parmentier2011strong,Mebrahtu2012quantum,AltimirasNoiseDCBPRL2014}, an effect known as Dynamical Coulomb Blockade (DCB) \cite{ingoldnazarov1992DCB}. By increasing the impedance of the electromagnetic environment to values comparable to the resistance quantum $R_{\mathrm{K}}=h/e^2\simeq 25.9\, \mathrm{k \Omega}$, the resulting strong coupling suppresses the transport at low voltage and low temperature for a normal conductor \cite{delsing1989,geerligs1989,ClelandDCBinTunnelPRB1992,parmentier2011strong,Mebrahtu2012quantum,AltimirasNoiseDCBPRL2014}. 

So far most DCB studies focused on the conductor's transport properties at low frequencies~\cite{buttiker2000,kindermann2003}, or  at higher frequencies~\cite{souquet2013,safi2014,grabert2015} but without describing photon radiation. Recent progress in microwave techniques opens new perspectives for investigating the quantum properties of the emitted radiation \cite{HofheinzBrightDCBPRL2011,bozyigit2011antibunching,GasseTunnelSqueezePRL2013,LiuDQDmaserScience2015}. Such radiative properties lie out of the scope of the standard DCB approach which focus primarily on electrons. Yet one may expect the emission of non-classical radiation in a strong coupling regime as DCB generally induces important non-linearities. In the case of a Josephson junction, it has indeed been predicted \cite{LeppakangasAntibunchedPhotonsPRL2015, AnkerholdJosepehsonSingleModePRB2015} that the emitted photons are strongly antibunched.

Focusing on photons rather than on electronic variables is also a well-suited point of view if one is interested in what is actually measured in a quantum circuit experiment at GHz frequencies. The classical and quantum backaction of the measurement channels are thus treated equally with the system. The question of modeling measurement was initially addressed by Lesovik and Loosen~\cite{LoosenLesovikJETP1997,GavishQNoisePRB2000} by coupling a quantum conductor to LC resonator representing the measurement apparatus. The general point of view considered here is that the system is ultimately connected to a transmission line carrying the photon radiation and that the measurements are realized on these output photons. Recently, the standard input-output theory of quantum optics~\cite{YurkeQNThyPRA1984,gardiner1985} has been adapted to describe the field response of quantum conductors~\cite{LeppäkangasNJOP2014,mendes2015,mendes2015}.

\begin{figure}[tb]
	\includegraphics[width=8.0cm]{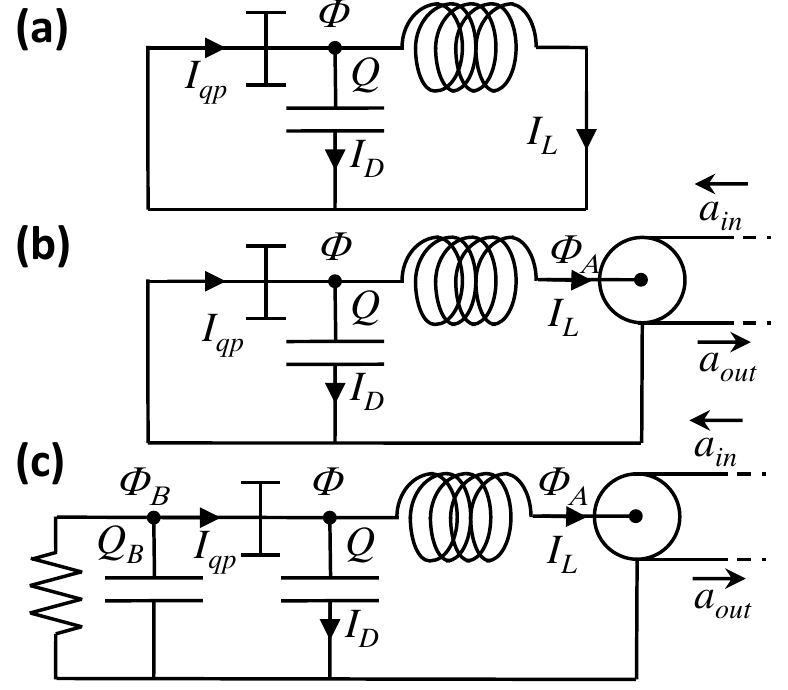}
\caption{Schematics of the considered circuits: (a) A tunnel junction with quasiparticle current $\hat{I}_{qp}$ is embedded in an $LC$ circuit described by conjugated fields at the depicted node: the inductive magnetic flux $\Phi$, and the capacitive influence charge $Q$. (b) A transmission line in series with the inductor damps the circuit by radiating outgoing modes $a_{out}$, which can can detected with a matched detection chain (not shown). (c) A high resistance ($R>R_{\mathrm{K}}$) $RC$ circuit  is connected to the junction. The resulting quantum flux fluctuations $\Phi_B(t)$ are responsible for strong dynamical Coulomb blockade modifying the quasiparticle current $\hat{I}_{qp}$, which can be efficiently collected in a matched detection band of the transmission line.}
\label{fig:circuits}
\end{figure}

 In this paper, we consider the case of a normal tunnel junction arbitrarily coupled to radiation. As a first step, we reconsider the model of Lesovik and Loosen and extend it to include measurement backaction, {\it i.e.} DCB induced by the LC resonator on the quantum conductor. We find that the energy transfer between the conductor and the resonator conserves the same structure as in the absence of DCB, a combination of emission and absorption noises, but with current correlators simply dressed by DCB inelastic processes. Next,  we develop a Hamiltonian approach which considers not only the electronic transport through the conductor but also the associated radiative dynamics via an input-output description \cite{YurkeQNThyPRA1984,gardiner1985}. Considering the Lesovik\&Loosen geometry complemented by a transmission line, we find that the energy transfer can be read out directly in the power of the output field. Finally, we exploit the Hamiltonian approach and propose a well-suited circuit geometry (Fig.~\ref{fig:circuits}(c)) in which a normal tunnel junction in the strong DCB regime can efficiently squeeze radiation. Squeezing in this scheme is induced by the dissipative bath provided by the tunnel junction~\cite{kronwald2013,didier2014}, together with direct parametric down-conversion resulting from the ac modulation of the field reflection coefficient.

The outline of the paper is the following: in Sec.~\ref{sec:lccircuit}, we first discuss the DCB effect of a high-impedance LC circuit on a tunnel junction, and evaluate the power emitted in the LC circuit. In Sec.~\ref{sec:dissip}, in order to account for energy dissipation in the system, we add a coupling to a dissipation line and set the stage for an input-output description of the junction-resonator circuit. In Sec.~\ref{sec:input}, extending the input-output analysis to a particular circuit where the tunnel junction is both strongly coupled to a DCB resistive circuit and weakly coupled to a transmission line for readout via a LC resonator, we show how strong squeezing emerges under a parametric ac driving.

\section{Power emitted with standard DCB} \label{sec:lccircuit}

As discussed in the introduction, the measurement of finite-frequency current fluctuation can be accounted for by the weak coupling to a LC resonator modeling the detector~\cite{LoosenLesovikJETP1997,GavishQNoisePRB2000}. The power emitted by the quantum conductor towards to the detector then accurately describes finite-frequency noise measurement. For a high-impendance and therefore strongly coupled LC resonator, the measurement backaction must be included in the formalism. We therefore consider a tunnel junction  element shunted by an $LC$ circuit of resonant frequency $\omega_{0}=1/\sqrt{LC}$ and characteristic impedance $Z_{LC}=\sqrt{L/C}$. We treat it in the standard DCB formulation \cite{ingoldnazarov1992DCB} assuming the LC detector to be always in a thermal state. Describing the relaxation dynamics of the detector will be the subject of the next section. 

 The electrodynamic coupling gives rise to inelastic tunneling events, modifying the charge transfer dynamics of the junction.  This physics is described by the Hamiltonian $H_0=H_{qp}+H_T+H_{LC}$, with 
\begin{equation}
H_{qp}=\sum_l \epsilon_l c_l^\dagger c_l+\sum_r \epsilon_r c_r^\dagger c_r
\end{equation}
 describing the left and right electrodes of the junction, 
\begin{equation}
H_{LC}=\frac{Q^2}{2C}+\frac{\Phi^2}{2L}
\end{equation}
the energy stored in the $LC$ circuit, and $H_T= T+T^\dagger$ with $T=\sum_{l, r}\tau_{l, r}c_l^\dagger c_r e^{ie\Phi/\hbar}$ the tunnel coupling which simultaneously transfers quasiparticles from the right to the left electrode with  amplitude $\tau_{l, r}$, while displacing the influence charge of the capacitance by the electron charge 
\begin{equation} e^{ie\Phi/\hbar}Qe^{-ie\Phi/\hbar}=Q-e,
\end{equation}
corresponding to the node commutation relation 
\begin{equation}
[\Phi,Q] = i \hbar.
\end{equation}
$H_T$ is the minimal coupling of the junction to its circuit, neglecting the intrinsic electrodynamics of the electrodes \cite{PierreNonDCBZBAPRL2001,LebedevPhotonStatsQPCPRB2010,cottet2015} beyond the mean-field approximation encompassed in the shunting capacitance $C$ and the inductance $L$. 

From this Hamiltonian we obtain the quasiparticle current 
\begin{equation}\label{eq:qpcurrent}
\hat{I}_{qp}=\mathrm{d}(e\sum_l (c_l^\dagger c_l))/\mathrm{d} t=\frac{ie}{\hbar}(T^\dagger-T), 
\end{equation}
the displacement current 
\begin{equation}
\hat{I}_D=\mathrm{d}{Q}/\mathrm{d} t=\frac{ie}{\hbar}(T^\dagger-T)-\Phi/L,
\end{equation}
and the inductive current $\hat{I}_{L}=\Phi/L$ which correctly compensate at the circuit node $\hat{I}_{qp}=\hat{I}_{D}+\hat{I}_{L}$ as required by gauge invariance. We now consider the radiative properties in the presence of both dc and ac bias, by evaluating the power emitted in the $LC$ circuit 
\begin{equation}
P_{LC}=\mathrm{d} H_{LC}/\mathrm{d} t=\frac{\hat{I}_{qp} Q + Q \hat{I}_{qp}}{2 C}
\end{equation}
and computing its expectation value to lowest order in the tunnel coupling $H_T$. To do so, we take the uncoupled boundary condition for the density matrix $\rho=\rho_{qp}\otimes\rho_{LC}$. The electrodes are initially at thermal equilibrium $\rho_{qp}=e^{-\beta H_{qp}}/Z_{qp}$. The $LC$ circuit is set in a displaced thermal state~\cite{FernColletSquezzedJMO1988,ParlavecchioFDTPRL2014} 
\begin{equation}
\rho_{LC}=D[\gamma]e^{-\beta H_{LC}}/Z_{LC}D^\dagger[\gamma]
\end{equation}
 where the displacement vector $\gamma=\frac{ieV_{ac}}{2 r \hbar \omega_0}$, with $r=\sqrt{\pi e^2 Z_{LC}/h},$ gives the deterministic voltage
\begin{equation}
\langle V(t)\rangle=Tr (\rho_{LC}\mathrm{d}\Phi/\mathrm{d} t)=V_{ac}\cos(\omega_0t) +V_{dc}.
\end{equation}
The average power reads \cite{Supplementary}:

\begin{align}\label{eq:Lesovik&Loosen}
\langle P_{LC} (t) \rangle =&\frac{(1+n_{B}(\hbar \omega_0))S_{I_{qp}}(\omega_0, t)-n_B(\hbar \omega_0)S_{I_{qp}}(-\omega_0, t)}{2C}\nonumber \\
&-\langle \hat{I}_{qp}(t) \rangle V_{ac}\cos(\omega_0 t),
\end{align}
where $n_B(\hbar \omega_0)$ is the bosonic thermal population of the $LC$ mode, 
\begin{equation}\label{powerspec}
S_{I_{qp}}(\omega, t)=\int \mathrm{d}\tau e^{-i \omega \tau}\langle \hat{I}_{qp}(t+\tau)\hat{I}_{qp}(t)\rangle
\end{equation}
is the power spectral density of quasiparticle current fluctuations \cite{ParlavecchioFDTPRL2014} (emission noise being here at positive frequency), and $\langle \hat{I}_{qp}(t)\rangle$ is the average quasiparticle current. In this expression both $S_{I_{qp}}(\omega_0, t)$ and $\langle I_{qp}(t)\rangle$ have an explicit time-dependence due to the breaking of time-translational invariance by the ac bias. 

The first term in Eq.~(\ref{eq:Lesovik&Loosen}) describes  the power being emitted/absorbed by the junction via its emission/absorption current fluctuations. We recover the same structure as for a weakly coupled LC detector in the absence of AC driving~\cite{LoosenLesovikJETP1997,GavishQNoisePRB2000},  the important difference being that the tunneling dynamics encompassed in $S_{I_{qp}}(\omega_0, t)$ take into account both DCB and photon-assisted tunneling effects. 
The second term describes the Joule power dissipated in the junction via its mean current response in phase with the ac excitation. Indeed, computing the power injected in the electrodes $P_{qp}=\mathrm{d}H_{qp}/\mathrm{d}t$ \cite{Supplementary} confirms that its average value is equal to the electrical power delivered by the dc source, minus that carried away by the $LC$ circuit which, here, acts both as a power source and sink: 
\begin{equation}
\overline{\langle P_{qp}(t)\rangle}=\overline{\langle \hat{I}_{qp}(t)\rangle V_{dc}}-\overline{\langle P_{LC}(t)\rangle}.
\end{equation}
This perturbative approach, valid in the high tunneling resistance limit, considers flux (voltage) fluctuations arising only from the external circuit dynamics. Moreover, it implicitly assumes the presence of additional mechanisms not specified in the Hamiltonian $H_0$, restoring the initial state of the full system in between every tunneling event. In the following we explicitly consider such mechanism.

\section{Circuit model for dissipation} \label{sec:dissip}

We now go beyond the standard DCB approach and include a dissipative channel in the model, see Fig.~\ref{fig:circuits}(b), by adding a semi-infinite transmission line~\cite{NyquistThm1928, YurkeQNThyPRA1984} characterized by the impedance $Z_{\ell}$. This not only provides (i) a precise mechanism for the damping of the LC circuit, but also (ii) a way to compute the properties of the radiation emitted by the junction into a linear detection circuit using an input-output approach. Our analysis thus extends previous works~\cite{leppakangas2014,grimsmo2016} by considering both quasiparticles and strong (DCB) backaction.  
We find in particular that the standard DCB formulation used to describe the circuit of Fig.~\ref{fig:circuits}(a) is justified in the limit case where the $LC$ resonator leaks photons in the transmission line much faster than it exchanges photons with the tunnel junction, so that a separation of time scales occurs. This corresponds to an impedance mismatch to the readout circuit $R_T \gg Z_{LC}^2/Z_{\ell}$, where $R_T$ denotes the junction's tunnel resistance and $Z_{\ell} = \sqrt{\ell / c} $ the transmission line characteristic impedance, with $\ell$ and $c$ standing respectively for the lineic inductance and capacitance

The energy stored in the LC mode now reads
\begin{equation}
H_{LC}=\frac{Q^2}{2C}+\frac{(\Phi - \Phi_A)^2}{2 L}.
\end{equation}
The dynamics of the transmission line is decribed by the Hamiltonian~\cite{YurkeQNThyPRA1984}:
\begin{equation}
H_{line} = \int_0^{+\infty} d x \left [ \frac{1}{2 \ell} \left( \frac{\partial \Phi_{\rm line} (x)}{\partial x} \right)^2 + \frac{q_{\rm line}(x)^2}{2 c} \right],
\end{equation}
where we introduced the conjugated variables  $\Phi_{\rm line} (x) $ and $q_{\rm line} (x) $ describing the flux and charge density in the line,  $x$ being the position along the line. The bosonic operators describing the input $a_{\rm in,\omega}$ and output $a_{\rm out,\omega}$ fields enter the mode decomposition of $\Phi_{\rm line} $,
\begin{equation}\label{eq:mode}
\Phi_{\rm line} (x) = \sqrt{\frac{\hbar Z_{\ell}}{8 \pi^2}} \int_0^{+\infty} \frac{d \omega}{\sqrt{\omega}} \left[ a_{\rm in,\omega} e^{-i k x} + a_{\rm out,\omega} e^{i k x} + {\rm h.c.} \right].
\end{equation}
The input-output theory is obtained by considering the time-evolution equations in the interaction representation and imposing the coupling between the line and the LC oscillator via  $\Phi_A \equiv \Phi_{\rm line} (0)$.

In the Heisenberg picture, the equations of motion  for the $LC$ circuit variables are $\partial_t \Phi = \partial_Q H = Q/C$ and $\partial_t Q =  - \partial_\Phi H$. Combined, they couple the fields of the line, the flux of the $LC$ circuit, 
\begin{equation}\label{eq:heisen1}
  C \partial_t^2 \Phi  = \frac{\Phi_A-\Phi}{L} + \hat{I}_{qp}^H , 
\end{equation}
and the current operator $\hat{I}_{qp}^H = e^{i H t} \hat{I}_{qp} e^{-i H t}$ defined in Eq.~\eqref{eq:qpcurrent} in the Heisenberg representation. No charge can accumulate on the node $A$, and we obtain a second equation
\begin{equation}\label{eq:heisen2}
 \frac{1}{\ell} \frac{\partial \Phi_{\rm  line} (0)}{\partial x}  =  \frac{\Phi_A-\Phi}{L},
\end{equation}
 from  $\partial_{\Phi_A} H = 0$. In a second step, the current operator is expanded in the linear response regime
\begin{equation}\label{eq:linear}
\hat{I}_{qp}^H (t) = \hat{I}_{qp} (t) + \frac{i}{\hbar} \int_{-\infty}^t d t' \, [H_T(t'),\hat{I}_{qp} (t)],
\end{equation}
where $\hat{I}_{qp} (t)$ denotes the current evolved with the Hamiltonian unperturbed by $H_T$.
Solving for equations~(\ref{eq:heisen1}) and~\eqref{eq:heisen2} in frequency  domain, one arrives at
\begin{equation}\label{result-inout}
a_{\rm out,\omega} = \frac{\Delta^* (\omega)}{\Delta (\omega)} a_{\rm in,\omega} - i \omega_0^2 \sqrt{\frac{2 Z_{\ell}}{\hbar \omega}} \, \frac{\hat{I}_{qp}^H (\omega)}{\Delta(\omega)},
\end{equation}
with $\Delta (\omega) = \omega^2 - \omega_0^2 + i \omega \kappa$. $\kappa = \frac{Z_{\ell}}{L}$ is  the LC resonator damping rate due to the transmission line. The first term corresponds to the input field reflected by the resonator with a phase shift (time delay). The second term is the field emitted by the tunnel junction itself and carrying current noise fluctuations. As such Eq.~\eqref{result-inout} does not fully solve the circuit dynamics since the output field still enters the current $\hat{I}_{qp}^H$ in the flux $\Phi$ dressing the tunneling operator $T$, calling for a self-consistent solution. 
%We shall therefore perform an approximation and neglect the quasiparticle current contribution to the fluctuations of $\Phi$.

However, writing the rescaled flux $\tilde{\Phi}(\omega) = \sqrt{2 \omega/(\hbar Z_{\ell} \omega_0^4)} \Phi(\omega)$ as 
\begin{equation}\label{phi1}
\tilde{\Phi}(\omega) = - \, \dfrac{2  a_{\rm in,\omega} + \left(\frac{i}{\omega}+\frac{1}{\kappa} \right) \sqrt{\frac{2 \omega Z_\ell}{\hbar}} \, \hat{I}_{qp}^H (\omega)}{\Delta(\omega)},
\end{equation}
we find that the flux fluctuations at frequency $\omega_0$ arising from the second term, which calls for the self-consistency, are negligible in the case of strong impedance mismatch 
\begin{equation}\label{eq:misatch}
\frac{S_{I_{qp}}(\omega_0) L^2 \omega_0}{\hbar Z_{\ell}} \sim \frac{Z_{LC}^2}{R_T Z_{\ell}} \ll 1.
\end{equation}
This can also be formulated in the time domain by inspecting the dynamics of the LC resonator: photons will leak much faster to the transmission line rather than to the tunnel junction~\cite{Souquet2014,mendes2015}, when $\kappa \gg (Z_{LC}/\hbar)[S_{I_{qp}}(-\omega_0)-S_{I_{qp}}(\omega_0)] \sim (Z_{LC}/R_T) \, \omega_0$, yielding the same small parameter as Eq.~\eqref{eq:misatch}. For circuits having this separation of time scales, the approximation $\hat{I}_{qp}^H=0$ in Eq.~\eqref{phi1} then reproduces standard DCB expressions~\cite{ingoldnazarov1992DCB,AltimirasNoiseDCBPRL2014} (see also appendix~\ref{app:DCB}) for phase fluctuations across the tunnel junction characterized by the impedance seen by the junction ${\rm Re} Z_t (\omega) = Z_{\ell} \omega_0^4/|\Delta (\omega)|^2$.  The resulting flux $\Phi$ is finally substituted in the current $\hat{I}_{qp}^H$ in Eq.~\eqref{result-inout}, enabling the calculation of spectral properties of the emitted light even for very strong DCB backaction. The net power carried out by the transmission line reads 
\begin{align}\label{eq:plt}
P_{TL} (t) &=  \langle A_{\rm out}^\dagger (t) A_{\rm out} (t)\rangle - \langle A_{\rm in}^\dagger (t) A_{\rm in} (t)\rangle,
\end{align}
where $A_{\rm in/out} (t) = \int_0^{+\infty} \frac{d \omega}{2 \pi} \sqrt{\hbar \omega} a_{\rm in/out,\omega} e^{-i \omega t}$. The details of its calculation are given in appendix~\ref{app:DCB} taking the incoming field to be described by a displaced thermal state.

As a result, in the high quality factor limit $\omega_0\gg \kappa$, the expression of $P_{TL} (t)$ agrees precisely with the mean power $\langle P_{LC} (t) \rangle$ derived in the previous section, see Eq.~\eqref{eq:Lesovik&Loosen}, with the same form as in the absence of DCB. It is worth noting that the products of the two terms in Eq.~\eqref{result-inout} mix the field and the junction dynamics giving rise to the Bose factors in Eq.~\eqref{eq:Lesovik&Loosen}, which vanish at low temperature $\hbar\omega_0\gg k_B T$, and to the Joule power dissipated in the junction. 

\section{Squeezed radiation} \label{sec:input}

We continue with the Hamiltonian approach and input-output framework to analyze a specific circuit, illustrated in Fig.~\ref{fig:circuits}(c), for which we will demonstrate that efficient squeezing in the output radiation can be realized.

 As strong DCB is responsible for non-linearities in transport, it is expected to also favor squeezing in the field emitted by the tunnel junction. The circuit of Fig.~\ref{fig:circuits}(b) is however not adapted to this effect. For strong impedance mismatch, the incoming mode is almost perfectly reflected by the junction, polluting the outgoing field with unsqueezed fluctuations. The impedance-matched junction~\cite{GasseTunnelSqueezePRL2013, grimsmo2016} on the other hand shunts environment fluctuations, thereby reducing non-linearities and squeezing efficiency.
We thus consider Fig.~\ref{fig:circuits}(c) where DCB and readout are spatially separated: on one side, the tunnel junction is coupled to a resistive circuit producing strong DCB, on the other side, a weakly coupled (i.e. low impedance) resonant circuit is used to probe the radiation emitted by the junction, providing a good impedance matching to the junction over a narrow bandwidth. In the following, we consider a situation where the classical bias at the junction consists in a dc voltage superimposed to an ac modulation at twice the resonator's frequency: $V_{\rm cl} (t)=V_{dc}+ V_{ac}\cos 2 \omega_0 t$. 

\subsection{Noise and linear response}

We assume the following hierarchy of resistances $R_T \gg R \gg R_{\mathrm{K}} \gg Z_{\ell},Z_{LC}$. The high resistance $R$ imposes strong fluctuations for $\Phi_B$ at the tunnel junction. An even larger $R_T$ is necessary to avoid shunting those fluctuations. In contrast to that, the resonant circuit produces weak flux fluctuations giving a negligible contribution to DCB effects and the flux $\Phi$ can be expanded to second order yielding the inductive coupling 
\begin{equation}\label{eq:hamilu}
H \simeq H^u  - \Phi  \, \hat{I}_{qp}- (e \Phi/\hbar)^2 H_T^u/2
\end{equation}
 in the Hamiltonian in which the tunnel coupling $T^u$ is dressed  by the flux $\Phi_B$ only,
\begin{equation}
H_T^u = T^u+T^{u \dagger}, \qquad T^u = \sum_{l, r}\tau_{l, r}c_l^\dagger c_r e^{-ie\Phi_B/\hbar}.
\end{equation}
 $H^u$ governs the uncoupled evolutions of the DCB tunnel junction and weakly damped LC resonator. Hence, dynamics of the tunnel junction and DCB resistive circuit have been isolated, only weakly probed by the readout circuit.

The input-output theory is constructed similarly to Ref.~\cite{grimsmo2016}: time evolution is still described by Eqs.~\eqref{eq:heisen1} and~\eqref{eq:heisen2}, the current operator is expanded in the flux $\Phi$ and in the linear reponse regime (tunnel limit) with Hamiltonian~\eqref{eq:hamilu},
\begin{equation}\label{eq:inter-pic}
\begin{split}
\hat{I}_{qp}^H (t) &= \hat{I}_{qp} (t) - \frac{i}{\hbar} \int_{-\infty}^t d t' \, \Phi (t') [ \hat{I}_{qp} (t'),\hat{I}_{qp} (t)] \\ 
&+ \frac{i}{\hbar} \Phi (t) \int_{-\infty}^t d t' \, \left(\frac{e}{\hbar}\right)^2 \left[  H_T^u (t') , H_T^u (t) \right].
\end{split}
\end{equation}
This double expansion is in fact justified by two small parameters: $Z_{LC}/R_K \ll 1$ ensures weak flux fluctuations for $\Phi$ and $R_K/R_T \ll 1$ controls the linear regime for the tunneling current. Eq.~\eqref{eq:inter-pic} can be rewritten in a more suggestive linear response form 
\begin{equation}\label{eq:linearresponse}
\hat{I}_{qp}^H (\omega) = \hat{I}_{qp} (\omega) - \sum_n Y_n (\omega) V(\omega- 2 n \omega_0)
\end{equation}
in the presence of the ac bias with frequency $2 \omega_0$, where we introduced the quantum voltage $V (t) = \dot{\Phi} (t)$. The admittances are related to the quasiparticles shot noise via a Kubo-like relation
\begin{equation}
Y_n (\omega) = i \int \frac{d \omega_1}{2 \pi} \frac{\delta {\cal S}_n (\omega_1) \, \omega }{\hbar (\omega-2 n \omega_0)(\omega-\omega_1^-) \omega_1^+}.
\end{equation}
Here $\omega_1^{\pm} = \omega_1 \pm i 0^+$ and $$\delta {\cal S}_n (\omega_1) = S_n(-\omega_1) - S_n(\omega_1- 2 n \omega_0),$$ where the quasiparticle current noise spectral power of Eq.~\eqref{powerspec} is expanded in Fourier components
\begin{equation}
S_{I_{qp}}(\omega, t)=  \sum_n S_n (\omega) e^{-2 i n \omega_0 t}.
\end{equation}
For a tunnel junction in the absence of DCB, the identity $S_n (-\omega_1) = S_n (\omega_1- 2 n \omega_0)$ for $n \ne0$ implies that all admittances $Y_{n \ne 0}$ vanish, indicating an absence of current rectification, as well as $Y_0 (\omega) = 1/R_T$, recovering the tunnel junction bare resistance. In the general case, the imaginary parts of the admittances $Y_n (\omega)$ all vanish for $\omega \to 0$ as there can be no phase shift with respect to the applied bias in the dc regime.

The evaluation of the functions $S_n(\omega)$, and thus $Y_n(\omega)$ is straightforward for a tunnel junction, using the standard $P(E)$ theory~\cite{safi2014,ParlavecchioFDTPRL2014}. The resulting expressions are given in appendix~\ref{app:Sn}. The effect of the very strong DCB assumed here is encoded in the function~\cite{ingoldnazarov1992DCB} 
\begin{equation}
P(E) = \frac{1}{\sqrt{4 \pi E_c k_B T}} e^{-(E-E_c)^2/(4 E_c k_B T)},
\end{equation}
which gives the probability of the environment to absorb an energy $E$ from a tunneling electron. $E_c = e^2/(2 C_B)$ is the charging energy. The equilibrium noise is expressed as a convolution product
\begin{equation}\label{eq:convo}
S_{eq} (\omega) = \int_{-\infty}^{+\infty} d E \, P(E) S_{eq}^{(0)} (\omega+E/\hbar),
\end{equation}
in terms of the finite-temperature noise power spectrum of a simple tunnel junction
\begin{equation}
S_{eq}^{(0)} (\omega) = \frac{2}{R_T} \frac{ \hbar \omega}{e^{\hbar \omega/k_B T}-1}.
\end{equation}

\subsection{Photon correlators}

The current response~\eqref{eq:linearresponse} is the missing piece needed to complete the input-output calculation with Eqs.~\eqref{eq:heisen1} and~\eqref{eq:heisen2}. Assuming $\kappa \ll \omega_0$, we obtain the boundary equation relating the input and output fields
%Details of the derivation are given in appendix~\ref{app:input}.} Assuming $\kappa \ll \omega_0$, Eq.~\eqref{eq:heisen} then leads in frequency space to
\begin{equation}\label{field-emitted2}
\begin{split}
&\left(\omega + i \kappa_+ \right) a_{\rm out,\omega+\omega_0} =  \left(\omega + i \kappa_- \right) a_{\rm in,\omega+\omega_0}  \\ & - i \sqrt{\frac{\omega_0 Z_{\ell}}{2 \hbar}} \hat{I}_{qp,\omega+\omega_0} 
- \frac{i Y_1}{2 C} \left( a_{\rm out,\omega_0-\omega}^\dagger - a_{\rm in,\omega_0-\omega}^\dagger \right)
\end{split}
\end{equation}
where $2 \kappa_{\pm} = Y_0/C \pm \kappa$, with the notation $Y_n \equiv Y_n (\omega_0)$. $Y_0/C$ and $\kappa$ are respectively the damping rates of the LC resonator to the tunnel junction and transmission line. 
One can check that in the absence of DCB, $Y_0 (\omega) = 1/R_T$ and $Y_1$ vanishes, so that only dissipative squeezing occurs~\cite{mendes2015,grimsmo2016}. On resonance and for impedance matched junction and resonator~\footnote{the condition of Eq.~\eqref{eq:match} can also be written as $\kappa = 1/(R_T C)$ in the absence of DCB. It corresponds to a perfect impedance matching of the tunnel junction to the transmission line through the LC circuit, or equivalently balanced damping rates~\cite{mendes2015} for photons in the LC resonator towards the resistive tunnel junction and the radiative transmission line.}
\begin{equation}\label{eq:match}
 \frac{Z_{LC}^2}{R_T Z_\ell} =  \frac{\mathcal{Q} Z_{LC}}{R_T} = 1,
\end{equation}
or $\kappa_-=0$, where $\mathcal{Q} = Z_{LC}/Z_{\ell}$ is the quality factor of the resonator, $a_{\rm out,\omega_0} \propto \hat{I}_{qp,\omega_0}$ and  squeezing properties in the noise fluctuations of the junction are imprinted in the output field~\cite{GasseTunnelSqueezePRL2013}. With DCB and higher non-linearities, $Y_1\ne 0$  and the last term in Eq.~\eqref{field-emitted2} introduces a parametric down-conversion mechanism on the input field, similarly to a parametric amplifier~\cite{Movshovich1990,castellanos2008}. In the general case, squeezing of the output field thus results from an interplay between these two mechanisms: squeezed radiation from the junction and direct parametric down-conversion. Note that the first r.h.s. term in Eq.~\eqref{field-emitted2} corresponds to the reflected part of the input field, detrimental to squeezing.

\begin{figure}
	\includegraphics[width=\columnwidth]{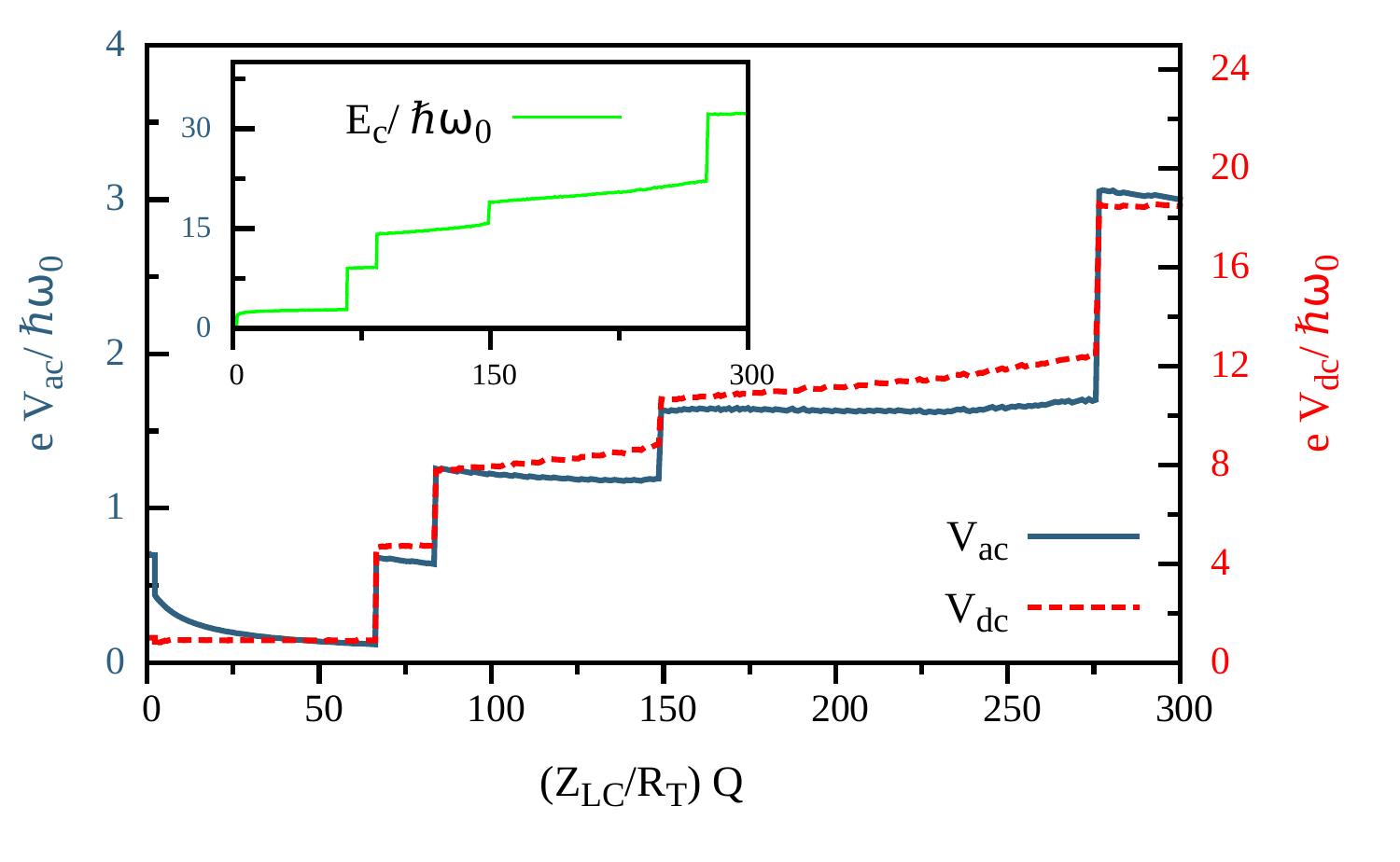}
	\includegraphics[width=0.5\textwidth]{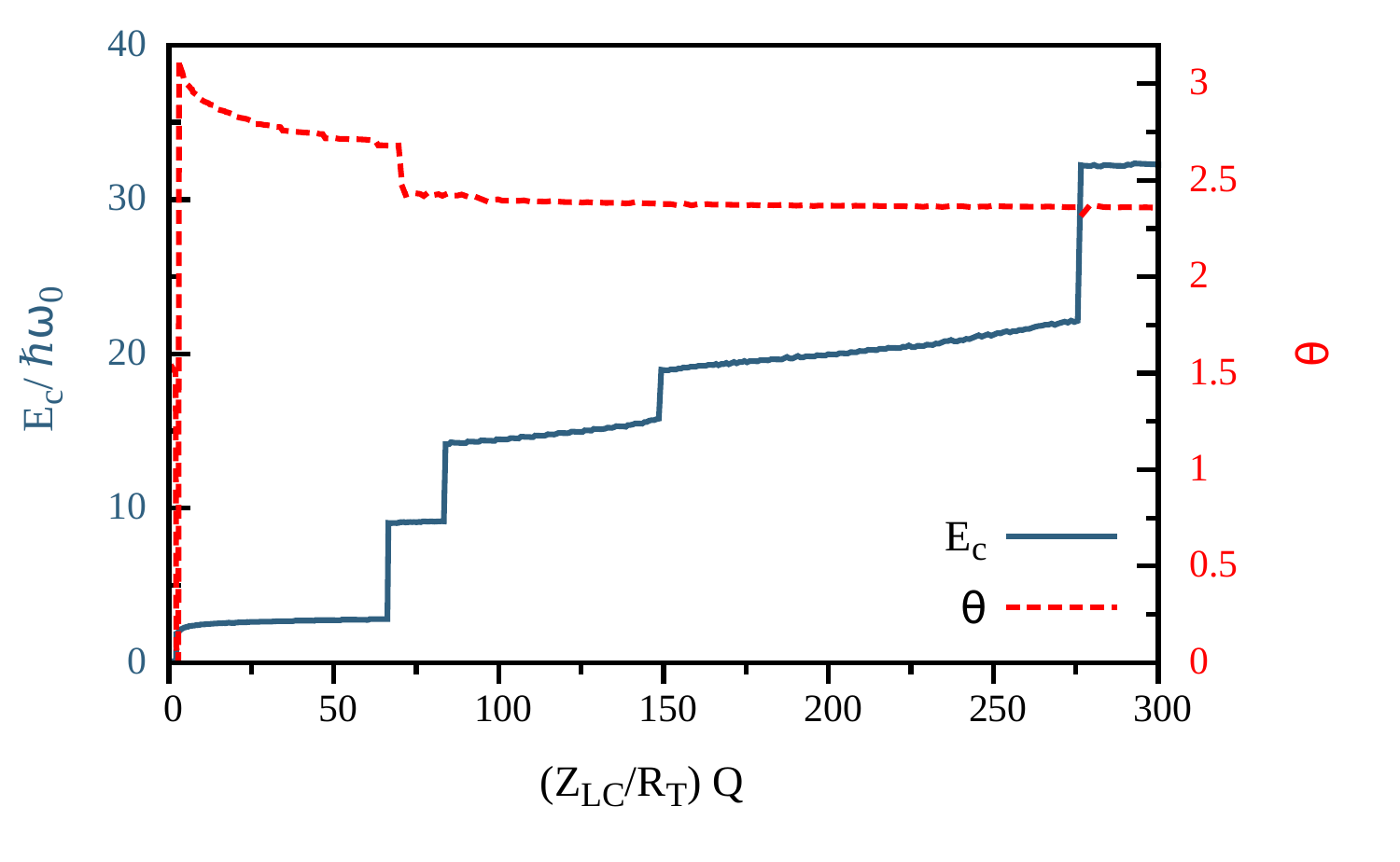}
\includegraphics[width=\columnwidth]{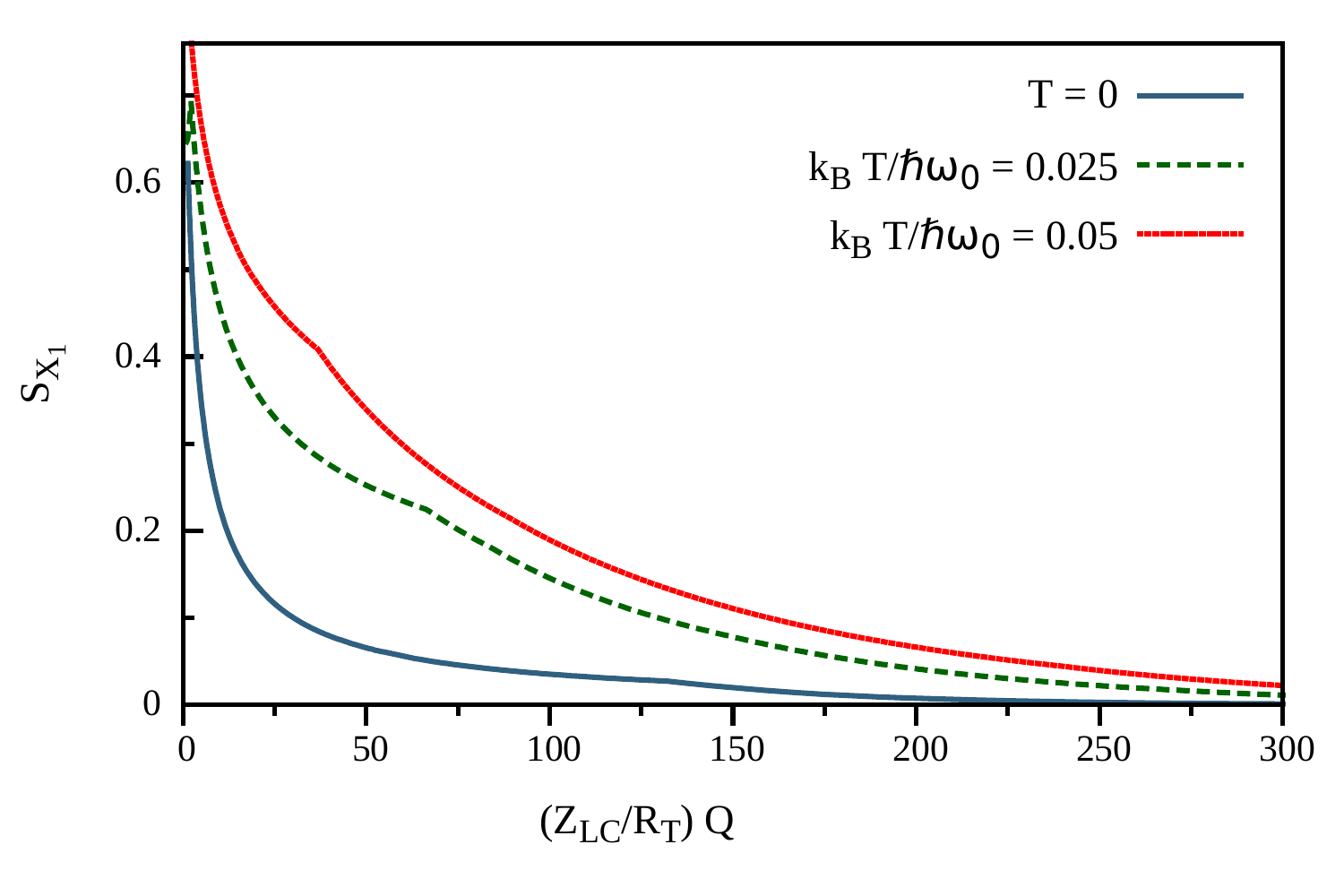}
\caption{Lower Panel: From Eq.~\eqref{field-emitted2}, squeezing of the output quadrature $X_{\theta,\omega} = e^{-i \theta} a_{\rm out,\omega+\omega_0} + e^{i \theta} a^\dagger_{\rm out,\omega_0-\omega}$, at the resonant frequency $\omega_0$, as function of the impedance-matching parameter $\mathcal{Q} Z_{LC}/R_T$ for very strong resistive DCB. The continuous line corresponds to the zero temperature limit, whereas the dashed and dotted curves correspond to temperatures of respectively 7 mK and 14 mK for a resonance at $\omega_0/2\pi$=~6~GHz. The charging energy $E_c$, the quadrature angle $\theta$ and the bias voltage at the junction are chosen to optimize (minimize) $S_{X_1}$ for each $\mathcal{Q} Z_{LC}/R_T$. The middle and upper panels give the associated variations of $E_C$, $\theta$, and of $V_{dc}$ and $V_{ac}$ at 7 mK.
\label{fig:squeezing}} 
\end{figure}

Eq.~\eqref{field-emitted2} is linear and can be easily inverted to express the output field $a_{\rm out,\omega_0+\omega}$ in terms of the fields $a_{\rm in,\omega_0+\omega}$,  $a_{\rm in,\omega_0-\omega}^\dagger$,  $\hat{I}_{qp,\omega+\omega_0}$ and $\hat{I}_{qp,\omega-\omega_0}$, in order to compute the output field correlations. The details of the calculation and the precise coefficients are given  in appendix~\ref{app:input}. Squeezing is characterized by the power spectrum $S_\theta (\omega)$ of the quadrature $X_{\theta,\omega} = e^{-i \theta} a_{\rm out,\omega+\omega_0} + e^{i \theta} a^\dagger_{\rm out,\omega_0-\omega}$, rotated by the angle $\theta$ compared to the quadrature in phase with the ac excitation, defined as
\begin{equation}\label{eq:quadra}
\langle \{ X_{\theta,\omega} , X_{\theta,\omega'} \} \rangle = 2 S_{\theta} (\omega) 2 \pi \delta (\omega + \omega')
\end{equation}
which can be measured with RF heterodyning scheme e.g.~\cite{LiuDQDmaserScience2015,ForguesTunnelPairsPRL2014}. For each value of $\mathcal{Q} Z_{LC}/R_T$, we determine numerically the optimal $\theta$, $E_c$, $V_{dc}$ and $V_{ac}$ which optimize squeezing. $S_\theta (\omega)$ is always minimum at $\omega=0$, with a bandwidth of the order of $\kappa$. Results for the squeezed quadrature at $\omega=0$, noted $S_{X_1}$, are displayed in Fig.\ref{fig:squeezing} at different temperatures. At $E_c/\hbar\omega_0=0$, only dissipative squeezing survives and we recover the values in the absence of DCB~\cite{GasseTunnelSqueezePRL2013,mendes2015,grimsmo2016}, namely $S_{X_1} = 0.618$ at zero temperature. Squeezing then improves with the ratio $\mathcal{Q} Z_{LC}/R_T$, showing that DCB effects can significantly improve the  squeezing efficiency.

At large $\mathcal{Q} Z_{LC}/R_T \gg 1$, the down-conversion mechanism dominates and a simple physical picture explains squeezing of the output field in analogy with flux-driven parametric amplifiers~\cite{johansson2009,wilson2011}. For strong DCB, the $2 \omega_0$ ac modulation of the classical bias drives the junction from a nearly insulating state with a very large impedance to a conducting state with an impedance near $R_T$. For $\mathcal{Q} Z_{LC} \gg R_T$, the corresponding microwave reflection coefficient between the LC circuit and the junction oscillates between its extremal values, thereby implementing a parametric drive of the resonator.
Microwave resonators with characteristic impedances in the range of a few kilo-Ohms have been reported using either kinetic  \cite{Altimirasimpedance13, PhysRevApplied.5.044004} or electromagnetic inductance \cite{2015arXiv151204660F} have been reported, with quality factor $\mathcal{Q}$ well exceeding $10^5$ (Ref.~\onlinecite{PhysRevApplied.5.044004}), allowing to reach values as high as a few $10^3$ for the ratio $\mathcal{Q} Z_{LC}/R_T$, while keeping $R_T$ in the 100 k$\Omega$ range to ensure that $R_T \gg R_K$. The optimum charging energy can be seen to increase with increasing $\mathcal{Q} Z_{LC}/R_T$, but remain within realistic boundaries:  Junctions with nanoscale cross-section \cite{JehlSEpumpPRX2013} can implement charging energies as large as $E_c/h=4\,\mathrm{THz}$. Promising squeezing levels, well above $10\,\mathrm{dB}$ thus seem within experimental reach.

In summary, we formulated a general input-output theory that captures at the same level strong dynamical Coulomb blockade physics and the quantum properties of the emitted light. We showed how strong blockade amplifies quadrature squeezing in the emitted field under parametric excitation. We gave specific results for the case of a tunnel junction but the generality of our approach makes it applicable to other conductors. The cases of quantum dots~\cite{FreyDpotCavityPRL2012,viennot2015coherent,petersson2012circuit} and hybrid systems such as SIS junctions~\cite{basset2010} seem particularly appealing for the purpose of squeezing efficiency. The extension of our approach to cotunneling processes, where two electrons may cooperate to emit a photon~\cite{tobiska2006,xu2014}, is another promising direction.

We gratefully acknowledge discussions with the Quantronics Group, P. Roche and G. Johansson. The research leading to these results has received funding from the European Research Council under the European Union's Programme for Research and Innovation (Horizon 2020) / ERC grant agreement number [639039], and support from the ANR AnPhoTeQ research contract.

%The opposite case where these two time scales are equal corresponds to an impedance matching situation where all photons incoming from the line are absorbed by the tunnel junction.

\appendix

\section{DCB power radiated in the transmission line}\label{app:DCB}

We consider the closed form~\eqref{result-inout} as the starting point for evaluating the output field correlations. Neglecting $\hat{I}_{qp}^H$ and using a thermal distribution for the input field, $\langle a_{\rm in,\omega}^\dagger a_{\rm in,\omega'} \rangle =  2 \pi  n_B (\hbar \omega)\delta (\omega-\omega')$, we recover standard DCB expressions~\cite{ingoldnazarov1992DCB,AltimirasNoiseDCBPRL2014} for phase fluctuations across the tunnel junction
\begin{subequations}
\begin{align}
& \langle e^{i e \Phi (t)/\hbar}  e^{-i e \Phi (0)/\hbar}  \rangle  \equiv e^{J(t)} \\[2mm]
\begin{split}
J(t) & = 2 \int_0^{+\infty} \frac{d \omega}{\omega} \frac{{\rm Re} Z_t (\omega)}{R_K} \times \\ 
& \times \left[ {\rm coth} \left( \frac{\beta \omega}{2} \right) ( \cos \omega t -1 ) - i \sin \omega t \right],
\end{split}
\end{align}
\end{subequations}
where $\beta = 1/(k_B T)$ is the inverse temperature of the input field and ${\rm Re} Z_t (\omega) = Z_0 \omega_0^4/|\Delta (\omega)|^2$ is the real part of the impedance seen by the junction.

Let us consider first the absence of an ac bias voltage. The power injected in the output field is expressed as
\begin{equation}\label{eq:power}
\langle  A_{\rm out}^\dagger (t)  A_{\rm out} (t) \rangle = \int_0^{+\infty} d\omega \, \hbar \omega f_{\rm out} (\omega) 
\end{equation}
where we introduced the photon-flux density from $\langle a_{\rm out,\omega}^\dagger a_{\rm out,\omega'}  \rangle = 2 \pi f_{\text{out}}(\omega) \delta(\omega - \omega')$. Defining the scattering phase $e^{i \theta_\omega} = \frac{\Delta^* (\omega)}{\Delta (\omega)}$ and the normalization factor ${\cal N}_\omega = \frac{\omega_0^2}{i \Delta (\omega)} \sqrt{\frac{2 Z_0}{\hbar \omega}}$, the decomposition Eq.~\eqref{result-inout} of the input field produces four terms in the calculation of $f_{\rm out} (\omega)$
\begin{equation}\label{decomp}
\begin{split}
\langle & a_{\rm out,\omega}^\dagger a_{\rm out,\omega'} \rangle = e^{i (\theta_{\omega'} - \theta_\omega)}  \langle a_{\rm in,\omega}^\dagger a_{\rm in,\omega'} \rangle \\[2mm]
& +  {\cal N}_{\omega'} e^{-i \theta_\omega} \langle a_{\rm in,\omega}^\dagger \hat{I}_{qp}^H  (\omega') \rangle +  {\cal N}_\omega^* e^{i \theta_{\omega'}} \langle  \hat{I}_{qp}^H  (-\omega)   a_{\rm in}(\omega') \rangle \\[2mm] & +  {\cal N}_\omega  {\cal N}_{\omega'}^* \langle \hat{I}_{qp}^H  (-\omega) \hat{I}_{qp}^H  (\omega')  \rangle.
\end{split}
\end{equation}
The first term is readily calculated
\begin{equation} 
e^{i (\theta_{\omega'} - \theta_\omega)}  \langle   a_{\rm in,\omega}^\dagger a_{\rm in,\omega'}    \rangle = 2 \pi \delta(\omega - \omega') n_B (\omega).
\end{equation}
It is equal to the photon-flux  $\langle  a_{\rm in,\omega}^\dagger a_{\rm in,\omega'}     \rangle$ of the incoming field $ a_{\rm in,\omega}$. This term is subtracted in the net output power $P_{LT}$ defined in Eq.~\eqref{eq:plt}.
The last term is written in terms of the power spectral density of quasiparticle current fluctuations $S_{I_{qp}} (\omega)$. It takes the form
\begin{equation}
{\cal N}_\omega  {\cal N}_{\omega'}^* \langle \hat{I}_{qp}^H  (-\omega) \hat{I}_{qp}^H  (\omega')  \rangle = | {\cal N}_\omega |^2 S_{I_{qp}} (\omega) 2 \pi \delta (\omega - \omega'),
\end{equation}
where only the first term is kept in the expansion~\eqref{eq:linear} of  $\hat{I}_{qp}^H$. If we use this result in the expression of the radiated power Eq.~\eqref{eq:power}, we arrive at the contribution $S_{I_{qp}} (\omega_0)/(2 C)$ under the assumption of a sharp resonance $\kappa \ll \omega_0$ and the integral
\begin{equation}\label{eq:inte}
\int_{0}^{+\infty} d \omega \, \hbar \omega  | {\cal N}_\omega |^2 \simeq \frac{Z_0 \omega_0^2}{2 \kappa} = \frac{1}{2 C},
\end{equation}
in agreement with the prefactor in Eq.~\eqref{eq:Lesovik&Loosen}.  $S_{I_{qp}} (\omega_0)$ is interpreted as the emission noise corresponding to the power emitted by current fluctuations in the tunnel junction, even in the absence of the input field. It takes into account the influence of DCB on transport and the noise can be written as a convolution  $S_{I_{qp}} (\omega) = \int d \varepsilon P(\varepsilon) S_{I_{qp}}^{0} (\omega-\varepsilon/\hbar)$ between the energy distribution function $P(E) = \frac{1}{h} \int_{-\infty}^{+\infty} d t \, e^{J(t) + i E t/\hbar}$ and the noise in absence of DCB effect
\begin{equation}
S_{I_{qp}}^0 (\omega) = \frac{1}{R_T} \sum_{\pm} \frac{\hbar \omega \pm e V_{dc}}{e^{\beta (\hbar \omega \pm e V_{dc})}-1}.
\end{equation}
The second and third terms in Eq.~\eqref{decomp} are complex conjugate to each other. In contrast to the last term in Eq.~\eqref{decomp}, 
it is now the second term in the expansion of Eq.~\eqref{eq:linear} which contributes to the calculation. The first term in Eq.~\eqref{eq:linear} creates electron-hole excitations across the junction and has a vanishing expectation. In the calculation, we make use of the following identity
\begin{equation}\label{identity}
\begin{split}
\langle a_{\rm in}^\dagger (t) [ H_T (t_1),& \hat{I}_{qp} (t_2) ] \rangle =  \langle [ \hat{I}_{qp} (t_1), \hat{I}_{qp} (t_2)]   \rangle \\ & \times  \left( \langle a_{\rm in}^\dagger  (t)  \Phi (t_2) \rangle - \langle a_{\rm in}^\dagger  (t) \Phi (t_1)  \rangle \right)  ,
\end{split}
\end{equation}
which is valid because the field  $a_{\rm in}$, and therefore $\Phi (\omega) = -i \hbar {\cal N}_\omega a_{\rm in} (\omega)$, have Gaussian distributions. After a tedious but straightforward calculation, we find the additional contribution $n_B (\omega) (S_{I_{qp}} (\omega) - S_{I_{qp}} (-\omega)) |{\cal N}_\omega |^2$ to $f_{\text{out}}(\omega)$. This term vanishes at zero temperature. We then integrate over frequencies using the integral Eq.~\eqref{eq:inte} and obtain the correction to the power in the output field
\begin{equation}
\frac{n_B (\omega) \left[S_{I_{qp}} (\omega) - S_{I_{qp}} (-\omega) \right]}{2 C}.
\end{equation}

To summarize, adding all contributions from Eq.~\eqref{decomp}, the calculation of the net output power $P_{LT}$ in the input-output formalism coincides with the power $\langle P_{LC} (t) \rangle$ received by the LC resonator in the standard DCB approach of Sec.~\ref{sec:lccircuit}.

The presence of an ac voltage can be included rigorously in the quantum formalism thanks to the displacement operator $D$ acting on both frequencies $\omega_0$ and $- \omega_0$. Quantum averages are then taken with respect to the displaced density operator $\rho = D e^{-\beta H} D^\dagger/Z$ and the action on the input field is given by
\begin{equation}
\begin{split}
D^\dagger a_{\rm in,\omega} D & =  a_{\rm in,\omega} + \frac{V_{ac}}{2 Z_{LC}} \sqrt{\frac{Z_0}{2 \hbar \omega_0}} \\ & \times \left[ 2 \pi \delta (\omega -\omega_0) - 2 \pi \delta (\omega +\omega_0) \right].
\end{split}
\end{equation}
Using the expression~\eqref{phi1} of $\Phi$, we obtain the shift in the flux
\begin{equation}
D^\dagger \Phi (\omega) D = \Phi (\omega) + \frac{i V_{ac}}{2 \omega_0} \left[ 2 \pi \delta (\omega -\omega_0) - 2 \pi \delta (\omega +\omega_0) \right],
\end{equation}
leading to $D^\dagger \frac{\partial \Phi (t)}{\partial t} D = \frac{\partial \Phi (t)}{\partial t} + V_{ac} \cos (\omega_0 t)$. Instead of dressing the density operator with $D$, it is possible to work with the undisplaced density operator $e^{-\beta H}/Z$ while all operators of the theory are dressed by $D$ and $D^\dagger$. The input-output relation~\eqref{result-inout} is then transformed to
\begin{equation}\label{result-inout-second}
\begin{split}
& a_{\rm out,\omega}  = \frac{\Delta^* (\omega)}{\Delta (\omega)} a_{\rm in,\omega} - \frac{V_{ac}}{2 Z_{LC}} \sqrt{\frac{Z_0}{2 \hbar \omega_0}} \\ & \times \left[ 2 \pi \delta (\omega -\omega_0) + 2 \pi \delta (\omega +\omega_0) \right] - i \omega_0^2 \sqrt{\frac{2 Z_0}{\hbar \omega}} \, \frac{\hat{I}_{qp}^H (\omega)}{\Delta(\omega)},
\end{split}
\end{equation}
where the flux in the current $\hat{I}_{qp}^H (t)$ contains the classical evolution $\frac{V_{ac}}{\omega_0} \sin(\omega_0 t)$. Inserting this result into the expression of the net injected power $P_{TL} (t)$ retrieves Eq.~\eqref{eq:Lesovik&Loosen} of Sec.~\ref{sec:lccircuit}.

\section{Current correlators}\label{app:Sn}

We consider a classical sinusoidal ac-bias applied to the tunnel junction $V_{\rm cl} (t)=V_{dc}+ V_{ac}\cos 2 \omega_0 t$.
The standard Fourier decomposition
\begin{equation}
e^{i \frac{e V_{ac}}{2 \hbar \omega_0} \sin (2 \omega_0 t)} = \sum_{m \in \mathbb{Z}} J_m \left( \frac{e V_{ac}}{2 \hbar \omega_0} \right) e^{2 i m  \omega_0 t}, 
\end{equation}
introducing the Bessel functions $J_n$, is used to derive the photo-assisted noise, or current-current correlators
\begin{equation}
\begin{split}
& S_n (\omega)  = \frac{1}{2} \sum_{m \in \mathbb{Z}} \left\{ J_m \left( \frac{e V_{ac}}{2 \hbar \omega_0} \right)  J_{m+n} \left( \frac{e V_{ac}}{2 \hbar \omega_0} \right) \right.  \\ & \times S_{eq} \left(\omega - e V_{dc}/\hbar -2 m \omega_0 \right)  
+  J_m \left( \frac{e V_{ac}}{2 \hbar \omega_0} \right)  \\ & \left. \times J_{m-n} \left( \frac{e V_{ac}}{2 \hbar \omega_0} \right) S_{eq} \left(\omega + e V_{dc}/\hbar +2 m \omega_0 \right) \right\}.
\end{split}
\end{equation}
In particular, the effect of DCB fluctuations factorizes and is entirely encoded in the equilibrium noise function $S_{eq} (\omega)$, see Eq.~\eqref{eq:convo}.

\section{Solution of the input-output equations}\label{app:input}

Injecting the linear response current expression~\eqref{eq:linearresponse} in Eqs.~\eqref{eq:heisen1} and~\eqref{eq:heisen2}, we arrive at the coupled equation
\begin{widetext}
\begin{align}
\nonumber &\bigg[\omega^2-\frac{1-Z_{\ell}  Y_0(\omega)}{LC}-i\omega\bigg(\frac{ Z_{\ell}}{L}-\frac{Y_0(\omega)}{C}\bigg)\bigg]a_{in,\omega} -\bigg[\omega^2-\frac{1+Z_{\ell} Y_0(\omega)}{LC}+i\omega\bigg(\frac{ Z_{\ell}}{L}+\frac{Y_0(\omega)}{C}\bigg)\bigg]a_{out,\omega}\\ \nonumber
&=\frac{i}{LC}\sqrt{\frac{2 Z_{\ell}}{\hbar\omega}}\hat{I}_{qp} (\omega)-\sum_{n^\ast<\frac{\omega}{2\omega_0}}\sqrt{1-\frac{2n\omega_0}{\omega}}Z_{\ell} Y_n(\omega)\bigg[\bigg(\frac{1}{LC}+i\frac{\omega-2n\omega_0}{Z_{\ell}C}\bigg)a_{in,\omega-2 n \omega_0}+\\ \nonumber
&\bigg(\frac{1}{LC}-i\frac{\omega-2n\omega_0}{Z_{\ell}C}\bigg)a_{out,\omega-2 n \omega_0} \bigg]+\sum_{n^\ast>\frac{\omega}{2\omega_0}}\sqrt{\frac{2n\omega_0}{\omega}-1}Z_{\ell} Y_n(\omega)\bigg[\bigg(\frac{1}{LC}-i\frac{2n\omega_0-\omega}{Z_{\ell}C}\bigg)a_{in,2 n \omega_0-\omega}^\dagger\\ \label{eq:monster}
&+\bigg(\frac{1}{LC}+i\frac{2n\omega_0-\omega}{Z_{\ell}C}\bigg)a_{out,2n\omega_0-\omega}^\dagger \bigg],
\end{align}
\end{widetext}
valid in the general case, where $n^\ast$ stands for $n\neq 0$. This lengthy expression can nevertheless be simplified in the limit of a high quality factor $\kappa \ll \omega_0$. First,  Eq.~\eqref{eq:heisen2} can be written  as $\Phi =  - \frac{L}{\ell} \partial_x \Phi_{\rm  line} (0) + \Phi_A$ where the second term is much smaller than the first one. Moreover, only the $n=0,1$ terms matter in Eq.~\eqref{eq:monster}, and other values of $n$ are suppressed in the limit $\kappa/\omega_0 \to 0$, filtered by the LC resonator. $n=1$ is the standard parametric term which couples frequencies $\omega_0$ and $-\omega_0$. After a few straigthforward algebraic manipulations assuming $\kappa \ll \omega_0$, we obtain the relation~\eqref{field-emitted2} given in the main text. It can also be written in matrix form,
\begin{equation}
\begin{split}
{\cal M}_+  \begin{pmatrix} a_{\rm out,\omega+\omega_0} \\ a^\dagger_{\rm out,\omega_0-\omega}
\end{pmatrix} & = {\cal M}_- \begin{pmatrix} a_{\rm in,\omega+\omega_0} \\ a^\dagger_{\rm in,\omega-\omega_0}
\end{pmatrix} \\
& - i \sqrt{\frac{\omega_0 Z_{\ell}}{2 \hbar}} \begin{pmatrix} \hat{I}_{qp,\omega+\omega_0}  \\ - \hat{I}_{qp,\omega-\omega_0}
\end{pmatrix} ,
\end{split}
\end{equation}
with
\begin{equation}
 {\cal M}_{\pm} = \begin{pmatrix} \omega + i \kappa_{\pm} & i Y_1/2 C \\  -i Y_1^*/2 C &  -\omega - i \kappa_{\pm}
\end{pmatrix}.
\end{equation}
The solution is 
\begin{equation}\label{sol-aout}
\begin{split}
a_{\rm out,\omega+\omega_0} & = \lambda_\omega \,  a_{\rm in,\omega+\omega_0} + \mu_\omega \, a_{\rm in,\omega_0-\omega}^\dagger 
\\ & + \alpha_\omega \, \hat{I}_{qp,\omega+\omega_0} + \beta_\omega \,  \hat{I}_{qp,\omega-\omega_0}, 
\end{split}
\end{equation}
with the frequency-dependent coefficients
\begin{equation}
\begin{split}
\lambda_\omega & = \frac{(\kappa_+ - i \omega) (\kappa_- -i \omega) - | Y_1/2 C |^2}{\cal D_\omega} \quad \mu_\omega = \frac{Y_1}{2 C} \frac{\kappa}{\cal D_\omega} \\[2mm] & \alpha_\omega = - \sqrt{\frac{\omega_0 Z_{\ell}}{2 \hbar}} \frac{\kappa_+ - i \omega}{\cal D_\omega}  \quad \beta_\omega =  \sqrt{\frac{\omega_0 Z_{\ell}}{2 \hbar}} \frac{Y_1}{2 C {\cal D}_\omega},
\end{split}
\end{equation}
and ${\cal D}_\omega = (\kappa_+-i \omega)^2 - | Y_1/2 C |^2$. They satisfy the unitary identity
$$|\lambda_\omega|^2 - |\mu_\omega|^2 + (|\alpha_\omega|^2 - |\beta_\omega|^2) \, 2 \hbar \omega_0 Y_0 = 1$$ to preserve the commutation relation of the output field $[a_{\rm out,\omega}, a^\dagger_{\rm out,\omega'}] = 2 \pi \delta(\omega-\omega')$.

The quadrature power spectrum $S_\theta (\omega)$ introduced in Eq.~\eqref{eq:quadra} indicates squeezing if $S_{\theta} (\omega) <1$. Using Eq.~\eqref{sol-aout}, we find $S_{\theta} (\omega) = A_1 (\omega) + e^{-2 i \theta} A_2 (\omega) + e^{2 i \theta} A_2^* (\omega)$ with
\begin{subequations}
\begin{align}
\begin{split}
A_1 & (\omega) = \left( |\lambda_\omega|^2 + |\mu_\omega|^2 \right) [1+2 n_B(\hbar \omega)] \\[2mm] & +  (|\alpha_\omega|^2 +  |\beta_\omega|^2) \left[ S_0 (\omega_0) +  S_0 (-\omega_0) \right] \\[2mm] & + 2 S_1 (\omega_0) {\rm Re} \left( \alpha_\omega^* \beta_\omega + \alpha_{-\omega}^* \beta_{-\omega} \right), 
\end{split} \\[2mm] 
\begin{split}
A_2 & (\omega) = (\lambda_\omega \mu_{-\omega} + \lambda_{-\omega} \mu_\omega ) [1/2+  n_B(\hbar \omega)] \\[2mm] & +\frac{1}{2} ( \alpha_\omega \beta_{-\omega} + \alpha_{-\omega} \beta_{\omega} ) \left[ S_0 (\omega_0) +  S_0 (-\omega_0) \right] \\[2mm] & + \left( |\alpha_\omega|^2 + \beta_\omega \beta_{-\omega} \right) S_1 (\omega_0). \label{eq:A2}
\end{split}
\end{align}
\end{subequations}
Using the polar representation $A_2 = |A_2| e^{i \varphi}$, one sees that the most squeezed quadrature is given by the angle $\theta = \varphi/2 + \pi/2$, and
\begin{equation}\label{eq:squeezing}
S_{\theta = \varphi /2 + \pi/2}  = A_1 - 2 |A_2|.
\end{equation}
This is in particular the angle chosen in Fig.~\ref{fig:squeezing}.

%\bibliography{MicroTunnelResub}

\newpage

\begin{widetext}

\setcounter{equation}{0}
\setcounter{section}{0}

\section*{Supplementary information for the article: Quantum Properties of the radiation emitted by a conductor in the Coulomb Blockade Regime}

These supplementary information provides the full calculations allowing us to derive the expressions present in the main article. Cited equations not preceded by $S-$ refer to the main text.
\section{Standard $P(E)$ approach}

\subsection{Electromagnetic power}

The power emitted into the $LC$ circuit is defined by the power operator $$P_{LC}=\mathrm{d}H_{LC}/\mathrm{d}t=\frac{1}{2C}\Big( \hat{I}_{qp}Q+Q\hat{I}_{qp}  \Big).$$

Here we compute the mean value of this operator up to lowest order in the coupling Hamiltonian $H_T$. Making use of the interaction picture of the power operator $P_{LC}^0(t)$ with respect to the uncoupled evolution $H_{qp}+H_{env}$, its time evolution up to first order in the tunnel coupling reads (here-after, the time-dependence of unlabeled operators are meant to be taken in the interaction picture):
\begin{align*}
P_{LC}^1(t)=&P_{LC}^0(t)+\frac{i}{\hbar}\int_{-\infty}^{0}[H_T(t+\tau),P_{LC}^0(t)]\mathrm{d}\tau
\end{align*}
Its quantum average over the initial states described in the article simplifies in:
\begin{align}\label{Eq:1stOrder}
\langle P_{LC}^1(t)\rangle=-\frac{1}{C}\operatorname{Re}\int_{-\infty}^0\frac{2e}{\hbar^2}\langle T(t+\tau)T^\dagger(t)Q(t)-T^\dagger(t+\tau)T(t)Q(t)\rangle -\langle \hat{I}_{qp}(t+\tau)\hat{I}_{qp}(t)\rangle \mathrm{d}\tau,
\end{align}
where we identified $\langle \hat{I}_{qp}(t+\tau) \hat{I}_{qp}(t)\rangle=\frac{e^2}{\hbar^2}\langle T(t+\tau)T^\dagger(t)+T^\dagger(t+\tau)T(t) \rangle$. Equation (\ref{Eq:1stOrder}) contains the real part of the already known quasiparticle current time-correlator, and two new correlation functions which we will now compute. Since the initial states are uncoupled, the correlation functions factorize in terms of quasiparticle and environment correlation functions. The quasiparticle correlation $\Theta(t+\tau)\Theta^\dagger(t)$, and $\Theta^\dagger(t+\tau)\Theta(t)$, where $\Theta=\sum_{l, r}\tau_{l,r}c_l^\dagger c_r$ is the quasiparticle tunneling operator, are already well known. In the case of a particle-hole symmetric systems (which is the case of metallic tunnel junctions probed in the relevant range of energies  much smaller than the barrier height and Fermi energy), one has:
\begin{align*}
\theta(\tau)=\langle\Theta(t+\tau)\Theta^\dagger(t)\rangle=\langle\Theta^\dagger(t+\tau)\Theta(t)\rangle=\frac{\hbar G_T}{2\pi e^2}\Bigg(i\pi\hbar\frac{d}{dt}\delta(\tau)-\frac{\pi^2}{\beta^2}\sinh^{-2}(\frac{\pi \tau}{\hbar \beta})\Bigg),
\end{align*}
where $G_T$ is the tunneling conductance, and $\beta$ the inverse temperature. Therefore, we only need to compute the environment correlation functions: $\langle e^{\pm i e \Phi(t+\tau)/\hbar}e^{\mp i e \Phi(t)/\hbar} Q(t) \rangle$. For the sake of clarity, we first compute them for a dc bias, and then discuss how an ac bias modifies this first result. 

\subsubsection{dc bias}

The magnetic flux operator reads: $\Phi(t)=V_{dc}t+\delta\Phi(t)$, with $\delta\Phi(t)=\frac{\hbar r}{e}(a(t)+a^\dagger(t))$, where $r=\sqrt{\frac{\pi Z_{LC}}{R_Q}}$ with $Z_{LC}=\sqrt{\frac{L}{C}}$ the mode impedance and $R_Q=h/e^2\simeq 25.8\,\mathrm{k}\Omega$ the resistance quantum, and where $a(t)$ and $a^\dagger(t)$ are correspondingly the mode annihilation and creation operators of the $LC$ in the interaction picture. Therefore the operator $e^{\pm ie\delta\Phi(t)/\hbar}$ and the products $e^{\pm ie\delta\Phi(t+\tau)/\hbar}e^{\mp ie\delta\Phi(t)/\hbar}$ can be recast with displacement operators $D[\alpha]=e^{\alpha a^\dagger-\alpha^\ast a}$:
\begin{align*}
&e^{\pm ie\delta\Phi(t)/\hbar}=D[\pm ire^{i \omega_0 t}],\\
&e^{\pm ie\delta\Phi(t+\tau)/\hbar}e^{\mp ie\delta\Phi(t)/\hbar}=e^{-ir^2\sin(\omega_0\tau)}D[\pm ire^{i\omega_0t}(e^{i\omega_0\tau}-1)].
\end{align*}
From this, we can compute the correlation function $e^{J(\tau)}=\langle e^{\pm ie\delta\Phi(t+\tau)/\hbar}e^{\mp ie\delta\Phi(t)/\hbar}\rangle$ appearing in $P(E)$ theory, being its inverse Fourier transform:
\begin{align*}
\langle e^{\pm ie\delta\Phi(t+\tau)/\hbar}e^{\mp ie\delta\Phi(t)/\hbar}\rangle&=\frac{e^{-ir^2\sin( \omega_0\tau)}e^{-\frac{1}{2}\beta \hbar\omega_0}e^{-r^2(1-\cos(\omega_0\tau))}}{Z_{LC}}\sum_n e^{-\beta n\hbar\omega_0}L_n^{0}(2r^2(1-\cos(\omega_0\tau)))\\
&=e^{J(\tau)}
\end{align*}
where $L_n^m(x)$ are generalized Laguerre polynomials of order $n$, from which we recover the well known expression for a single mode
\begin{equation*}
J(\tau)=r^2\bigg( \big(\cos(\omega_0\tau)-1\big)\coth\Big(\frac{\beta \hbar\omega_0}{2}\Big)-i\sin(\omega_0\tau)  \bigg).
\end{equation*}

A similar calculation gives the result:
\begin{align}\label{Eq:newCorr}
\langle&e^{\pm ie\Phi(t+\tau)/\hbar}e^{\mp ie\Phi(t)/\hbar} Q(t) \rangle=\pm\frac{e}{2}e^{\pm ieV_{dc}\tau/\hbar}\Bigg( \big(1-e^{i\omega_0\tau}+e^{\beta \hbar\omega_0}e^{-i\omega_0 \tau}\big)e^{J(\tau)}-\frac{ie^{\beta \hbar \omega_0}\big(e^{J(\tau)}\big)'}{\omega_0 r^2}  \Bigg).
\end{align}

Inserting Eq. (\ref{Eq:newCorr}) back in Eq. (\ref{Eq:1stOrder}), together with the already known expression for quasiparticle current fluctuations $ \langle \hat{I}_{qp}(t+\tau)\hat{I}_{qp}(t)\rangle=\frac{2e^2}{\hbar^2}\theta(\tau)e^{J(\tau)}\cos(eV_{dc}\tau/\hbar)$,  the average electromagnetic power reads:
\begin{align}\label{Eq:LesovikDC}
\langle P_{LC}^1(eV_{dc}) \rangle=&-\frac{2e^2}{\hbar^2 C}\operatorname{Re}\int_{-\infty}^0 \cos(eV_{dc}\tau/\hbar)\theta(\tau)\Bigg[\big(e^{\beta \hbar \omega_0}e^{-i\omega_0 \tau}-e^{i\omega_0\tau}\big)e^{J(\tau)}-\frac{ie^{\beta \hbar\omega_0}\big(e^{J(\tau)}\big)'}{\omega_0 r^2}  \Bigg)\Bigg] \mathrm{d}\tau\nonumber\\ 
&=\frac{1}{2C}\Big( \big( 1+n_B(\hbar\omega_0) \big)S_{I_{qp}}(V_{dc},\omega_0)-n_B(\hbar\omega_0)S_{I_{qp}}(V_{dc},-\omega_0)\Big)
\end{align}
which is the article Equation (1) specialized to the case of a dc bias. It can also take the following form:
$$ \langle P_{LC}^1(eV_{dc}) \rangle= \int_{-\infty}^{+\infty} \frac{S_V(-\omega)S_{I_{qp}}(V_{dc},\omega)}{\hbar\omega}\frac{\mathrm{d}\omega}{2\pi},$$
where we introduced the spectral density of voltage fluctuations of the $LC$ circuit: 
$$ S_V(\omega)=\int_{-\infty}^{+\infty} \mathrm{d}\tau e^{-i\omega \tau} \langle \dot{\Phi}(t+\tau)\dot{\Phi}(t) \rangle - \langle \dot{\Phi}(t+\tau)\rangle\langle\dot{\Phi}(t) \rangle.$$
This expression is more symmetric in the sense that the power emitted (absorbed) from the tunnel junction via its current fluctuations is proportional to the spectral density of emission (absorption) current fluctuations of the junction multiplied by the spectral density of absorption (emission) voltage fluctuations of the load electromagnetic environment. Note however that this expression is not fully symmetric: while the voltage fluctuations appear via their closed cumulant, the current fluctuations appear via their raw moment.

\subsubsection{ac bias}
We now take the boundary condition described in the article, $$\rho_{LC}(t\to -\infty)=D[\gamma]\frac{e^{-\beta H_{LC}}}{Z_{LC}}D^{\dagger}[\gamma],$$
describing a thermal field being displaced by a "classical" source. The displacement vector $\gamma= i V_{ac} \sqrt{C/(2 \hbar \omega_0)}$ gives rise to a deterministic time-dependent ac voltage: $\langle V(t)\rangle=Tr\big( \rho_{LC}(t\to -\infty) \dot{\Phi}(t)\big)=V_{dc}+V_{ac}\cos(\omega_0 t)$, without perturbing the quantum and thermal voltage fluctuations. Again, we exploit the properties of displacement operators in order to get:
\begin{align*}
\langle \hat{I}_{qp}(t+\tau)\hat{I}_{qp}(t)\rangle=\frac{2e^2}{\hbar^2}\theta(\tau)e^{J(\tau)}\cos\Big(\frac{eV_{dc}\tau}{\hbar}+\frac{eV_{ac}}{\hbar\omega_0}\big(\sin(\omega_0(t+\tau)-\sin(\omega_0 t)\big)\Big)
\end{align*}
and
\begin{align*}
&\langle e^{ie\Phi(t+\tau)/\hbar}e^{-ie\Phi(t)/\hbar}Q(t)-e^{-ie\Phi(t+\tau)/\hbar}e^{ie\Phi(t)/\hbar}Q(t) \rangle=\\
&e \cos\Big(\frac{eV_{dc}\tau}{\hbar}+\frac{eV_{ac}}{\hbar\omega_0}\big(\sin(\omega_0(t+\tau)-\sin(\omega_0 t)\big)\Big)\Bigg( \big(1-e^{i\omega_0\tau}+e^{\beta \hbar\omega_0}e^{-i\omega_0 \tau}\big)e^{J(\tau)}-\frac{ie^{\beta \hbar\omega_0}\big(e^{J(\tau)}\big)'}{\omega_0 r^2}  \Bigg)\\
&+2i\sin\Big(\frac{eV_{dc}\tau}{\hbar}+\frac{eV_{ac}}{\hbar\omega_0}\big(\sin(\omega_0(t+\tau)-\sin(\omega_0 t)\big)\Big)\theta(\tau)e^{J(\tau)}CV_{ac}\cos(\omega_0 t),
\end{align*}
From which we obtain the first equation of the article:
\begin{align*}
\langle P_{LC}^1(t)\rangle&=\frac{1}{2C}\Big( \big(1+n_B(\hbar\omega_0)\big) S_I(\omega_0, t) - n_B(\hbar\omega_0)S_I(-\omega_0, t) \Big)-\langle \hat{I}_{qp}^1(t)\rangle  V_{ac}\cos(\omega_0 t)\\
&=\int_{-\infty}^{+\infty} \frac{S_V(-\omega)S_{I_{qp}}(\omega, t)}{\hbar\omega}\frac{\mathrm{d}\omega}{2\pi}-\langle \hat{I}_{qp}^1(t)\rangle  V_{ac}\cos(\omega_0 t).
\end{align*}
Contrary to the stationary case, where power is exchanged only via the current and voltage fluctuations of the circuit, now the junction can also dissipate some energy initially contained in the $LC$ circuit via the average time-dependent current response, as stressed by the last equality.

\subsection{Joule power}

We define now the power injected within the electrodes:
\begin{align*}
P_{qp}=\frac{i}{\hbar}[H_{0}, H_{qp}]= -\frac{i}{\hbar}\sum_{l, r}(\epsilon_l-\epsilon_r)\tau_{l, r}c_l^\dagger c_r e^{ie\Phi/\hbar}+(\epsilon_r-\epsilon_l)\tau^\ast_{l, r}c_r^\dagger c_l e^{-ie\Phi/\hbar},
\end{align*}
and we expand its time evolution to first order in the tunnel coupling:
\begin{align*}
P_{qp}^1(t)=P_{qp}^0(t)+\frac{2}{\hbar^2}\operatorname{Re}\int_{-\infty}^0 \sum_{l,r}&T^\dagger(t+\tau) (\epsilon_l-\epsilon_r)\tau_{l, r}c^\dagger_l c_r e^{i(\epsilon_l-\epsilon_r)t/\hbar} e^{ie\Phi(t)/\hbar}\\
&-T(t+\tau)(\epsilon_l-\epsilon_r)\tau^\ast_{l, r}c^\dagger_r c_l e^{-i(\epsilon_l-\epsilon_r)t/\hbar} e^{-ie\Phi(t)/\hbar}\mathrm{d}\tau.
\end{align*}
Again, the evaluation of this operator with the uncoupled boundary conditions factorizes in quasiparticle and environment correlation functions. The environment correlation functions are just the standard $
\langle e^{\pm ie\Phi(t+\tau)/\hbar}e^{\mp ie\Phi(t)/\hbar}\rangle=e^{\pm i \big( \frac{eV_{dc}\tau}{\hbar}+\frac{eV_{ac}}{\hbar\omega_0}(\sin(\omega_0(t+\tau))-\sin(\omega_0t))\big)}e^{J(\tau)}.$ Then we only need to compute the new quasiparticle correlation functions:
\begin{align*}
\langle \Theta(t+\tau)\sum_{l,r}(\epsilon_l-\epsilon_r)\tau_{l, r}^\ast c_r^\dagger c_l e^{-i(\epsilon_l-\epsilon_r)t/\hbar}\rangle &= -i\hbar \dot{\theta}(\tau),\\
&= -\langle \Theta^\dagger (t+\tau)\sum_{l,r}(\epsilon_l-\epsilon_r)\tau_{l, r} c_l^\dagger c_r e^{i(\epsilon_l-\epsilon_r)t/\hbar}\rangle
\end{align*}
where the last equality holds for the particle-hole symmetric junctions.

Picking up the terms we obtain for a dc bias:
$$\langle P_{qp}^1(eV_{dc})\rangle=\frac{2\pi}{\hbar}\Big(  \epsilon\theta(\epsilon)\ast P(\epsilon)_{(eV_{dc})}+ \epsilon\theta(\epsilon)\ast P(\epsilon)_{(-eV_{dc})}\Big),$$
while for an ac bias we get:
\begin{align*}
\langle &P_{qp}^1(t)\rangle=\frac{2}{\hbar^2}\operatorname{Re}\int_{-\infty}^0\mathrm{d}\tau 2\cos(\frac{eV_{dc}\tau}{\hbar}+\frac{eV_{ac}}{\hbar\omega_0}(\sin(\omega_0(t+\tau))-\sin(\omega_0t)))i\hbar\dot{\theta}(\tau)e^{J(\tau)}.\\
\end{align*}

\subsection{Power balances}
\subsubsection{dc bias}
Since there is neither a dc voltage drop across the inductance, nor a dc displacement current, the average electrical power which is supplied by the voltage source is directly:
\begin{align*}
\langle \hat{I}_{qp}(eV_{dc}) \rangle V_{dc}&=\frac{2\pi}{\hbar}\int \mathrm{d}E (eV_{dc}-E) \theta(E)P(eV_{dc}-E) +(-eV_{dc}-E) \theta(E)P(-eV_{dc}-E)\\
&+\frac{2\pi}{\hbar}\int \mathrm{d}E E \theta(E)P(eV_{dc}-E) +E\theta(E)P(-eV_{dc}-E),
\end{align*}
where we identify the power absorbed in the quasiparticles \begin{align*}
P_{qp}(eV_{dc})=\frac{2\pi}{\hbar}\int \mathrm{d}E E \theta(E)P(eV_{dc}-E) +E\theta(E)P(-eV_{dc}-E).
\end{align*}
The other term can be worked out to match the average power emitted into the environment:
\begin{align*}
\frac{2\pi}{\hbar}\int \mathrm{d}E (eV_{dc}-E) \theta(E)P(eV_{dc}-E) +(-eV_{dc}-E) \theta(E)P(-eV_{dc}-E)=P_{LC}^1(eV_{dc})
\end{align*}

With these identifications we obtain the stationary power balance: $$P_{DC}(eV_{dc})=P_{qp}(eV_{dc})+P_{LC}(eV_{dc}).$$

\subsubsection{ac bias}

In the presence of the ac bias, the average power delivered by the dc source reads:
$$\overline{ \langle \hat{I}_{qp}(t)\rangle V_{dc}}=\sum_k J_k^2\Big( \frac{eV_{ac}}{\hbar\omega_0} \Big) I_{qp}(eV_{dc}+k\hbar\omega_0)V_{dc},$$
where we exploited the Jacobi-Angers expansion of exponentials having trigonometric arguments. On the other hand, the average dissipative ac response reads:
\begin{align*}\overline{ \langle \hat{I}_{qp}(t)\rangle V_{ac}\cos(\omega_0 t)}&=\sum_k J_k\Big( \frac{eV_{ac}}{\hbar\omega_0} \Big)\frac{V_{ac}}{2}\Big(J_{k+1}\Big( \frac{eV_{ac}}{\hbar\omega_0} \Big)+J_{k-1}\Big( \frac{eV_{ac}}{\hbar\omega_0} \Big)\Big) I_{qp}(eV_{dc}+k\hbar\omega_0)\\
&=\sum_k \frac{k\hbar\omega_0}{e}J_k^2\Big( \frac{eV_{ac}}{\hbar\omega_0} \Big) I_{qp}(eV_{dc}+k\hbar\omega_0).
\end{align*}
Combining the two expression with the results for a dc bias, we get the power balance of the circuit:
$$\overline{ \langle \hat{I}_{qp}(t)\rangle V_{dc}}=\overline{P_{qp}(t)} +\overline{P_{LC}(t)}.$$

\end{widetext}

\end{document}